\documentclass[preprint,superscriptaddress,nofootinbib]{revtex4-1}
\usepackage{hyperref}
\usepackage{bbm}
\usepackage{amsfonts}
\usepackage{mathrsfs}

\usepackage{amsmath,amssymb}

\usepackage{epsfig}
\usepackage{graphicx}               % Standard graphics package
\usepackage{url}
\usepackage{hyperref}
\usepackage{float}
\usepackage{pstricks}
\usepackage{color}

\newcommand{\nc}{\newcommand}

\nc{\beq}{\begin{equation}}
\nc{\eeq}{\end{equation}}
\nc{\beqa}{\begin{eqnarray}}
\nc{\eeqa}{\end{eqnarray}}
\nc{\bea}{\begin{eqnarray}}
\nc{\eea}{\end{eqnarray}}
\nc{\barray}{\begin{eqnarray}}
\nc{\earray}{\end{eqnarray}}
\nc{\barrayn}{\begin{eqnarray*}}
\nc{\earrayn}{\end{eqnarray*}}
\nc{\ra}{\rightarrow}
\newcommand{\lsim}{\!\mathrel{\hbox{\rlap{\lower.55ex \hbox{$\sim$}} \kern-.34em \raise.4ex \hbox{$<$}}}}
\newcommand{\gsim}{\!\mathrel{\hbox{\rlap{\lower.55ex \hbox{$\sim$}} \kern-.34em \raise.4ex \hbox{$>$}}}}
\nc{\Tr}{{\rm Tr}}
\nc{\slsh}{\slash\hspace*{-0.22cm}}

\def\be{\begin{equation}}
\def\ee{\end{equation}}
\def\bea{\begin{eqnarray}}
\def\eea{\end{eqnarray}}
\def\bit{\begin{itemize}}
\def\eit{\end{itemize}}

\nc{\infinity}{\infty}
\nc{\mc}{\mathcal}
\nc{\M}{\mathcal{M}}

\newcommand{\pv}{\langle\phi\rangle}
\newcommand{\hv}{\langle H_u^0\rangle}

\def\AMS{{\sc Ams-02}}

\begin{document}

\title{~\\PeV Neutrinos and a 3.5 keV X-Ray Line\\ from a PeV Scale Supersymmetric Neutrino Sector}

\author{\vskip 0.5cm Samuel B. Roland, Bibhushan Shakya, James D. Wells \vskip 0.05cm \textit{Michigan Center for Theoretical Physics, \\University of Michigan, Ann Arbor, MI 48109, USA \\ ~}}

\begin{abstract}
Recent measurements of PeV energy neutrinos at IceCube and a 3.5 keV X-ray line in the spectra of several galaxies are both tantalizing signatures of new physics. This paper shows that one or both of these observations can be explained within an extended supersymmetric neutrino sector. Obtaining light active neutrino masses as well as phenomenologically interesting (keV-GeV) sterile neutrino masses without any unnaturally small parameters hints at a new symmetry in the neutrino sector that is broken at the PeV scale, presumably tied to supersymmetry breaking.  The same symmetry and structure can sufficiently stabilize an additional PeV particle, produce its abundance through the freeze-in mechanism, and lead to decays that can give the energetic neutrinos observed by IceCube. The lightest sterile neutrino, if at 7 keV, is a non-resonantly produced fraction of dark matter, and can account for the 3.5 keV X-ray line. The two signals could therefore be the first probes of an extended supersymmetric neutrino sector.
\end{abstract}

\preprint{MCTP-15-13}
\preprint{CETUP2015-009}

\maketitle

%\tableofcontents

%%%%%%%%%%%%%%%%%%%%%%%%%%%%%%%%
\section{Introduction and Motivation}
\label{sec:introduction}
%%%%%%%%%%%%%%%%%%%%%%%%%%%%%%%%

Astrophysical signatures of dark matter (DM) interactions, coupled with theoretical and observational input from other areas of particle physics, can provide crucial insights into the nature of physics beyond the Standard Model (BSM). However, while the existence and macroscopic properties of dark matter in our Universe have been well established, its microscopic properties continue to remain elusive. Compelling theoretical arguments dictate that dark matter is most likely composed of weakly interacting massive particles (WIMPs), predicting a GeV-TeV mass scale for dark matter, although supporting observational evidence has so far failed to materialize. %On the contrary, null results at numerous indirect and direct detection experiments and the first run of the LHC seem to require a departure from the WIMP paradigm as well as traditionally studied models of new physics at the weak scale.
Interestingly, some recent observations hint at dark matter beyond this narrow GeV-TeV window, indicating that the structure of dark matter and BSM physics might be very different from what was envisioned.

One such observation is the IceCube neutrino observatory's detection of 37 neutrino events in the energy range from 30 GeV to 2 PeV, disfavoring a purely atmospheric explanation at $5.7 \sigma$ \cite{Aartsen:2014gkd}. In particular, IceCube has reported 3 neutrino events above 1 PeV. Strongly incompatible with traditional astrophysical processes, these have been shown to be compatible with the decay of heavy, long lived particles with lifetimes of about 10$^{27}$s that constitute some or all of dark matter \cite{Fong:2014bsa, Ema:2013nda, Ema:2014ufa, Feldstein:2013kka, Esmaili:2013gha, Esmaili:2014rma, Bai:2013nga, Bhattacharya:2014vwa,Higaki:2014dwa,Bhattacharya:2014yha,Rott:2014kfa}. A dark matter interpretation of these events, however, poses several theoretical inconveniences: Why should we expect new physics at the PeV scale? Why should a PeV scale particle be so long-lived? What sets its relic density? Why should it show up in neutrinos? Given the dearth of BSM candidates at the PeV scale, a theoretically motivated framework that also fits into the broader particle physics picture would be extremely appealing.

Another observation, made independently by two groups, is the discovery of an unidentified X-ray line at $E_\gamma \approx 3.5$ keV in the stacked X-ray spectra of 73 galaxy clusters measured by XMM-Newton \cite{Bulbul:2014sua} and in the X-ray spectra of the Andromeda galaxy and the Perseus galaxy cluster \cite{Boyarsky:2014jta}. A dark matter interpretation of these observations is plagued by many unresolved questions, such as possible contamination from potassium lines \cite{Jeltema:2014qfa,Boyarsky:2014paa, Bulbul:2014ala} as well as inconsistency with stacked observations of dwarf spheroidals \cite{Malyshev:2014xqa} and galaxy spectra \cite{Anderson:2014tza} (although greater compatibility is reported with observations of the Milky Way \cite{Riemer-Sorensen:2014yda, Boyarsky:2014ska}). Despite these concerns, we entertain the possibility that these observations can be explained by a monochromatic line from the decay of a 7 keV dark matter particle. Unlike the IceCube PeV neutrinos, there exists an extremely well-motivated keV scale dark matter candidate in the form of a sterile neutrino, whose decay into an active neutrino and an X-ray photon with energy equal to half its mass has long been heralded as its smoking gun signature. Sterile neutrinos are essential ingredients of seesaw models of neutrino masses, and while their masses could, a-priori from theory, lie anywhere from the eV to the GUT scale, forming a significant fraction of dark matter while remaining consistent with X-ray \cite{Boyarsky:2006fg,Boyarsky:2006ag, Watson:2006qb,Boyarsky:2005us,Boyarsky:2007ay,Boyarsky:2007ge,Watson:2011dw,Horiuchi:2013noa} and Lyman-alpha \cite{Abazajian:2005xn,Seljak:2006qw, Asaka:2006nq, Viel:2006kd,Boyarsky:2008xj,Polisensky:2010rw,Viel:2013fqw} measurements constrains them to be at the keV scale.

The purpose of this paper is to show that one or both of these signals can naturally arise from an extended supersymmetric neutrino sector. The framework is a simple extension of a recently proposed model \cite{Roland:2014vba}, where it was shown that if the right-handed neutrinos are charged under some new symmetry $U(1)'$, broken by the PeV scale vacuum expectation value (vev) of a scalar field, active neutrino masses consistent with data and keV-GeV mass sterile neutrinos that can form part or all of dark matter can be realized without the need for any unnaturally small parameters in the theory. The PeV scale vev can presumably be inspired by the scale of supersymmetry breaking. There exist compelling arguments for supersymmetry at such high scales from flavor, CP, and unification considerations \cite{Wells:2003tf, ArkaniHamed:2004fb, Giudice:2004tc, Wells:2004di}. High scale SUSY is now further motivated by null results at numerous indirect and direct detection experiments and the first run of the LHC. The Higgs boson mass $m_h=125$ GeV is also compatible with PeV scale sfermions for $\mathcal{O}$(1) values of tan$\beta$ in the MSSM \cite{Giudice:2011cg, ArkaniHamed:2012gw, Arvanitaki:2012ps}.

The realization of the 3.5 keV line from a sterile neutrino component of dark matter in this framework is straightforward. In addition, we will see that a straightforward extension of the neutrino sector using the same $U(1)'$ symmetry and structure employed for neutrino masses can result in a PeV scale dark matter candidate whose decays are compatible with the high energy neutrino events at IceCube. It is interesting to note that a mixture of cold and warm dark matter components might also be favorable for solving some of the small scale problems in cosmology \cite{Lovell:2011rd, BoylanKolchin:2011dk}.

Finally, a few words on the philosophy behind the structure paper. This paper does not aim to build a complete model of PeV scale supersymmetry -- this would bring in model-dependent details and complexities that are irrelevant to the task at hand. We simply intend to demonstrate that our framework can naturally accommodate the IceCube PeV neutrinos and the 3.5 keV X-ray line; for this reason, we also do not pursue detailed scans over parameters, and the benchmark values presented in this paper should not be interpreted as the ``best-fit" ones. The greatest virtue of this exercise lies not in obtaining best fits or building a complete model but in placing the two signals in perspective within a broader particle physics framework, where they are not isolated observational curiosities but are connected to outstanding issues in the neutrino and Higgs sectors. Finally, although unifying such disparate scales as the keV and PeV into a single simple framework appears extremely appealing, it should be kept in mind that these two signals are by no means necessary ingredients in the theory; indeed, both, one, or neither of them can be realized in the model with appropriate choices of fields and parameters.

%%%%%%%%%%%%%%%%%%%%%%%%%%%%%%%%
\section{The Model}
\label{sec:model}
%%%%%%%%%%%%%%%%%%%%%%%%%%%%%%%%

In this section we present our model of the supersymmetric neutrino sector, which aims to realize active neutrino masses consistent with oscillation data, a keV sterile neutrino dark matter candidate, and a PeV scale dark matter candidate without any unnaturally small parameters in the theory. This is an extension of the model in \cite{Roland:2014vba}, and the interested reader is referred to this paper for further details. As motivated in \cite{Roland:2014vba}, we take the scale of supersymmetry breaking to be around $\mathcal{O}(1-100)$ PeV. This means that all dimensionful parameters obtained after supersymmetry breaking, such as masses and vevs, are expected to be at this scale.

In order to obtain neutrino masses, three SM-singlet sterile neutrinos $N_i$ are introduced. Although the $N_i$ are singlets under the SM gauge group, they are unlikely to be singlets under all symmetries of nature, as this would naturally place their masses at the Planck scale or the GUT scale, contrary to what is phenomenologically desirable. We therefore posit that the $N_i$ are charged under some new symmetry of nature (for concreteness, a $U(1)'$). %This new symmetry also forbids the traditional terms of the seesaw mechanism, which would otherwise require extremely small Yukawa couplings $\sim 10^{-7}$ in order to obtain a keV scale sterile neutrino dark matter candidate.
The $N_i$ are coupled to a single exotic field $\phi$ of equal and opposite $U(1)'$ charge to form $U(1)'$ singlets, which can then be coupled to SM fields.

With the IceCube PeV neutrinos in mind, we also introduce a new field $X$, a SM singlet scalar with a PeV scale mass and appropriately charged under the $U(1)'$ to be sufficiently long-lived to form a component of dark matter. It is likewise coupled to another field $Y$ (which carries lepton number) to form a $U(1)'$ singlet. Given the new symmetry $U(1)'$, this is the most straightforward extension of the model to incorporate additional fields.

\begin{table}
\centering
\begin{tabular}{ | c | c | c | c | l |}
\hline
~Supermultiplet~ & ~spin 0, 1/2~ & $~U(1)'~$ &~ Remarks \\
\hline
  $\mathcal{N}_i$ & $\tilde{N_i}$, $N_i$ & +1 &  ~$N_i$ sterile neutrinos ~\\
  $\Phi$ & $\phi, \psi_\phi$ & -1 &  ~$\pv\sim$PeV, breaks $U(1)'$ ~\\
  $\mathcal{X}$ & X, $\psi_X$ & +5 &  ~$m_X\sim$PeV, dark matter~ \\
  $\mathcal{Y}$ & Y, $\psi_Y$ & -5 &  ~ $U(1)'$ partner of $\mathcal{X}$~\\
  \hline
\end{tabular}
\caption{Field content, notation, and $U(1)'$ charge assignments for the new multiplets introduced in the neutrino sector of the model. These lead to the higher-dimensional operators in the superpotential in Eq.\,\ref{eq:superpot}.}
\label{table:charges}
\end{table}

Since the theory is supersymmetric, each of these fields resides in a chiral supermultiplet; the field content and notation are summarized in Table \ref{table:charges}. These lead to the following non-renormalizable terms in the superpotential that are relevant to our study:
\beqa
\label{eq:superpot}
{\mathcal{W}}\supset && ~\frac{\zeta_{ij}}{M_*} L_i H_u \,\mathcal{N}_j \Phi+\frac{\alpha_i}{M_*} L_i H_u \mathcal{X} \mathcal{Y}\,+\frac{\eta_i}{M_*}\mathcal{N}_i \mathcal{N}_i \Phi\Phi+\frac{\lambda_1}{M_*}\mathcal{X X Y Y}+\frac{\beta_i}{M_*}\mathcal{N}_i \Phi \mathcal{X Y}\,\nonumber\\
&& ~ +\frac{1}{5!}\frac{\lambda_2}{M_*^3} \mathcal{X} \Phi^5\,+\frac{1}{5!}\frac{\lambda_3}{M_*^3} \mathcal{Y N}_i^5.
\eeqa
%We have only listed the terms that are directly relevant for this paper, dropping terms such as $H_u\,H_d\, \mathcal{N}_i \Phi$ and $\mathcal{Y} \mathcal{N}\,^5$ that are also generated but do not play any meaningful role in this study.
All couplings are written as dimensionless numbers and expected to be $\mathcal{O}(1)$. The $N_i$ basis is chosen such that the third term in Eq.\,\ref{eq:superpot} is diagonal. $M_*$ is the scale at which this effective theory of non-renormalizable operators needs to be UV completed with new physics, such as the scale of grand unification $M_{GUT}$ or the Planck scale $M_P$.

Obtaining Dirac and Majorana masses for the sterile neutrinos $N_i$ in order to recover the seesaw mechanism requires the scalar component $\phi$ of $\Phi$ to obtain a vev, thereby breaking the $U(1)'$ symmetry. We assume that $\phi$ obtains a PeV scale vev from the supersymmetry breaking sector (without delving into details of how exactly this might be realized, which is tangential to the main purpose of this paper). In addition, we also assume that the fields in the $\mathcal{X,Y}$ and $\Phi$ multiplets all get PeV scale masses. This setup has the following phenomenological consequences:

\subsection{Neutrino Masses}

With $\phi$ obtaining a vev at the PeV scale and $H_u$ acquiring a vev from electroweak symmetry breaking, the first and third terms in the superpotential in Eq.\,\ref{eq:superpot} lead to the following active-sterile Dirac masses and sterile Majorana masses in the neutrino sector (flavor indices suppressed for simplicity):
\beq
m_D=\frac{\zeta\pv\hv}{M_*},~~~~~m_M=\frac{\eta\pv^2}{M_*}.
\eeq
The seesaw mechanism then gives the following sterile and active neutrino mass scales:
\be
\label{Mas}
m_s = m_M=\frac{\eta \pv^2}{M_*}, ~~~m_a = \frac{m_D^2}{m_M}=\frac{\zeta^2 \hv^2}{\eta M_*},
\ee
which also determines the mixing angle between the active and sterile sectors:
\be
\label{eq:mixing}
\theta \approx \sqrt{\frac{m_a}{m_s}} = \frac{\zeta \hv}{\eta \pv}.
\ee

With $\pv\! \sim\!  1\! -\! 100$\,PeV, $M_*\!=\!M_{GUT}\,=\!10^{16}\, {\rm GeV}$, tan$\beta\! =\!2\,(\hv\!=\!155.6$\,GeV), and $\mathcal{O}$(1) values of $\zeta$ and $\eta$, this framework produces active neutrino masses that fit oscillation data and sterile neutrinos with ${\mathcal{O}}$(keV-GeV) masses, which are compatible with dark matter and cosmological observations (see \cite{Roland:2014vba} for more details).

\subsection{Sterile Neutrinos and Dark Matter}
\label{sterileneutrino}

The three sterile neutrinos $N_i$ naturally have masses at the keV-GeV scale in this framework. We require the lightest one, $N_1$, to be a dark matter candidate with a keV scale mass in order to explain the 3.5 keV X-ray signal. However, several recombination era observables \cite{Kusenko:2009up, Hernandez:2014fha, Vincent:2014rja,Vincent:2014rja} constrain the two heavier sterile neutrinos $N_2, N_3$ to decay before Big Bang Nucleosynthesis (BBN), forcing $\tau_{N2,N3}\lesssim 1$s and consequently $m_{N2,N3}\gtrsim\mathcal{O}(100)$ MeV. This mass hierarchy between $m_{N1}$ and $m_{N2,3}$ requires a similar hierarchy between $\zeta_{ij}$ and $\eta_i$ values, necessitating some tuning of parameters.

As $N_1$ couples extremely weakly to the SM fields and is never in thermal equilibrium in the early Universe, its relic abundance is not set by thermal freeze-out. It is produced instead through active-sterile oscillation at low temperatures, known as the Dodelson-Widrow (DW) mechanism \cite{Dodelson:1993je}. A combination of X-ray bounds and Lyman-alpha forest data now rule out the prospect of all of dark matter consisting of $N_1$ produced via the DW mechanism; however, it can still constitute a significant fraction of the dark matter abundance (this will be further discussed in Section \ref{sec:xray}). This is desirable for us, since the remaining dark matter can come from the PeV sector.

\subsection{PeV Scale Dark Matter}
\label{sec:scalarx}

Our PeV dark matter candidate is a SM singlet scalar $X$ that carries a $U(1)'$ charge (see Table\,\ref{table:charges}). In general, a PeV scale dark matter candidate presents two caveats. First, obtaining the correct relic density through the well-known thermal freeze-out mechanism requires a large annihilation cross-section that comes into conflict with unitarity limits \cite{Griest:1989wd}, so that freeze out of a PeV scale particle overcloses the Universe. Hence one has to ensure that the candidate is never in thermal equilibrium and build up its abundance through some other mechanism. Second, since the decay rate of a particle is in general proportional to its mass, an unstable PeV scale particle is generically far too short-lived to be a dark matter candidate, and appropriate measures need to be put in place to stabilize it against rapid decay. We will see that the new $U(1)'$ symmetry in our theory can be used to address both of these issues.

In light of the unitarity bound mentioned above, some further assumptions need to be made about the superpartners that freeze out of the thermal bath, as we take supersymmetry to be at the PeV scale. One possibility is to make the lightest supersymmetric particle (LSP) sufficiently light that the thermal freeze-out abundance does not overclose the Universe. Another is to have all superpartners decay into SM states through R-parity violating (RPV) interactions. In this paper we choose the former option despite the clear need to tune parameters in order to achieve this, as the latter would involve the introduction of several RPV operators in our model, which require careful treatment beyond the scope of this work. To this end, the LSP is chosen to be a Higgsino at $\sim800$\,GeV, which would make up about half of the dark matter abundance.

\vskip 0.5cm
\noindent\textit{1.~~~Production of X}
\vskip 0.3cm

%\vskip -0.5cm
%\subsubsection{Production of X}
Since $X$ is charged under the $U(1)'$, there are no renormalizable terms in the superpotential that connect it with SM fields. The lowest dimension term allowed, which must be both a SM singlet and a $U(1)'$ singlet, is the term $\frac{\alpha_i}{M_*}L_i H_u \mathcal{X} \mathcal{Y}$ (see Eq.\,\ref{eq:superpot}); this leads to the following production processes for $X$ from the thermal bath:
\begin{equation}
l\,h\,\rightarrow X \,\psi_Y,~~~~~ l\,\tilde{H}\,\rightarrow X \,Y,~~~~~\tilde{l}\,\tilde{H}\,\rightarrow X\,\psi_Y.
\label{xproduction}
\end{equation}
Here $l$ denotes both charged leptons and neutrinos, and $h$ denotes both neutral and charged higgses, and likewise for their superpartners $\tilde{l}$ and $\tilde{H}$. The above processes are suppressed by $M_*$ and therefore not strong enough to bring $X$ into equilibrium. Rather, since these interactions are extremely feeble, the abundance of $X$ gradually builds up via the process of freeze-in \cite{Hall:2009bx} as long as the processes remain kinematically feasible. Given the nonrenormalizable operators that leads to these interactions, the interaction cross sections scale as $\sim s/M_*^2$, where $s$ is the center of mass energy of the annihilation process, and the production rate is proportional to the temperature of the Universe, being the greatest at the earliest times.

The same processes also result in freeze-in abundances of $Y,\, \psi_X$, and $\psi_Y$. Assuming $m_{\psi_X}\,\textgreater\, m_Y\,\textgreater\, m_{\tilde{H}}, m_X$ and $m_{\psi_Y}\,\textgreater\,m_X$, these particles then decay via
\begin{equation}
\psi_X\,\rightarrow Y\,l\,h,~~~~~ Y\,\rightarrow X\,l\,\tilde{H},~~~~~\psi_Y\,\rightarrow X\,l\,h,
\label{exoticdecays}
\end{equation}
converting the abundances of $Y$, $\psi_Y$, and $\phi_X$ into $X$ abundance. Taking all these contributions into account, we calculate the relic abundance of $X$ to be approximately (in agreement with previously derived results in \cite{Elahi:2014fsa, Kusenko:2010ik, Blennow:2013jba, Khalil:2008kp})
\be
\Omega_{X}h^2 \sim 0.12 ~\bigg(\frac{m_{X}}{10 ~\text{PeV}}\bigg) \bigg(\frac{\alpha}{10^{-4}} \bigg)^2  \left(\frac{T_{RH}}{1.5 \times 10^{10} ~\text{GeV}} \right) \left(\frac{10^{16} ~\text{GeV}}{M_*} \right)^2
\ee
where we have taken $\alpha=\alpha_i$ for simplicity. Therefore, with a sufficiently high reheat temperature $T_{RH}$ and appropriate values of $\alpha$, the PeV scale particle $X$ could compose a significant fraction of dark matter.

\vskip 0.5cm
\noindent\textit{2.~~~Decay of X}
\vskip 0.3cm

Next, we must ensure that $X$ has a lifetime much longer than the age of the Universe and the correct decay rate and channels to produce the neutrinos observed at IceCube. We have already chosen $m_{\psi Y}\,,m_Y\,\textgreater \,m_X$, hence the only term in the superpotential Eq.\,\ref{eq:superpot} that can cause $X$ to decay is $\frac{1}{5!}\frac{\lambda_2}{M_*^3} \mathcal{X} \Phi^5\,$. Assuming $\langle \phi\rangle\textgreater \,m_\phi$, the leading decay process is $X\rightarrow \psi_\phi\,\psi_\phi$, coming from the Lagrangian term
\begin{equation}
\mathcal{L} \supset -\frac{\lambda_2}{12} \left(\frac{\pv}{M_*}\right)^3 X \, \psi_\phi \, \psi_\phi\,.
\end{equation}
Here we assume, for simplicity, that the decay $X\rightarrow \phi\phi$ or the decays from mixing with $\phi$ induced by the corresponding soft term $\frac{A_X}{M_*^3}X\phi^5$ that appears after supersymmetry breaking are subdominant (however, we have checked that these channels also give similar neutrino spectra to that from $X\rightarrow \psi_\phi\,\psi_\phi$). Assuming $m_{\psi_\phi}/m_X\ll 1$, this decay process has a lifetime
\begin{equation}
\label{eq:lifetime}
\tau_X\approx 10^{27} \,\text{s}\, \left(\frac{1.5\times10^{-3}}{\lambda_2}\right)^2\,\left(\frac{M_*}{10^{16}\, \text{GeV}} \right)^6 \, \left(\frac{100\, \text{PeV}}{\pv} \right)^6 \, \left(\frac{\text{PeV}}{m_X} \right).
\end{equation}

The $\psi_\phi$ further decays as $\psi_\phi\rightarrow N\tilde{H}\nu\,,\psi_\phi\rightarrow N\tilde{H^\pm}l^\mp$ through an off-shell sterile sneutrino as a consequence of the $L_i H_u \,\mathcal{N}_j \Phi$ and $\mathcal{N}_i \mathcal{N}_i \Phi\Phi$ terms in the superpotential. The sterile neutrinos $N$ then further decay through the standard sterile neutrino decay channels to produce additional active neutrinos.

As the decay lifetime required to fit the IceCube data is $\tau\sim 10^{27}$\,s, Eq.\,\ref{eq:lifetime} suggests that one can obtain the necessary lifetime for reasonable choices of parameters in the model (see section \ref{sec:icecube} below). Note the role of the $1/M_*^3$ suppression in obtaining such a long lifetime; this was the motivation behind the choice of the $U(1)'$ charge of $+5$ for $\mathcal{X}$.

%%%%%%%%%%%%%%%%%%%%%%%%%%%%%%%%
\section{Compatibility with Signals}

In this section we demonstrate the compatibility of the IceCube neutrino and 3.5 keV X-ray line signals with the framework described in the previous section. As mentioned in the introduction, one could incorporate neither, one, or both of these into the model with appropriate parameter choices. In this section we choose to include both, as a proof of principle that both can be incorporated simultaneously into the framework. To demonstrate this, we work with a specific choice of parameters, which are listed in Table \ref{table:parameters}; the active and sterile neutrino masses and relic abundances of various dark matter components that result from these choices are also listed. As stressed in the Introduction, these are not best-fit points resulting from some scan but simply a judicious choice of parameters to achieve the desired results.

\begin{table}[t]
\begin{ruledtabular}
\begin{tabular}{cccc}
%$\pv=110$ PeV & $M_*=M_{\rm GUT} ~= 10^{16}$ GeV & $\tan\beta=2$ & $T_{RH}=10^{10}$ GeV \\
%\hline\hline
\multicolumn{1}{l}{\underline{Couplings}} &~&~&~\\
 \underline{$\zeta_{ij}$} & \underline{$\eta_i$} &   &  \\
 $\left(
\begin{array}{ccc}
3.53 & -2.28 & -1.19 \times 10^{-5}\\
1.02 & -3.54 & -1.99 \times 10^{-5} \\
-0.65 & -1.28 & 3.38 \times 10^{-5} \\
\end{array}
\right)$ & $\begin{array}{c} 5.82 \times 10^{-6}\\1.26\\1.67\\ \end{array}$ & $\begin{array}{c} \alpha=0.007\\ \lambda_2=0.0002 \\ \\ \end{array}$ & \\
\hline
\multicolumn{1}{l}{\underline{Masses}} &~&~&~\\
\underline{$m_a$ (eV)} & \underline{$m_s$} & ~ & ~ \\
$\begin{array}{c}7.75\times 10^{-7}\\0.0087\\0.049\end{array}$ & $\begin{array}{c} 7.00 ~\text{keV}\\ 1.50~\text{GeV} \\2.00 ~\text{GeV}\end{array}$ & $\begin{array}{c} m_X=7~ \text{PeV}\\m_{\psi_\phi}=2 ~\text{PeV}\\m_{\tilde{H}^0,\tilde{H}^\pm}=800 ~\text{GeV}\end{array}$ & ~\\
\hline
\multicolumn{1}{l}{\underline{Dark Matter Properties}} &~&~&~\\
$m_{N_1} = 7$ keV & $m_X=7$ PeV & $m_{\tilde{H}^0}=800$ GeV & ~\\
$\Omega_{N_1}h^2=0.03~(=25\%)$ & $\Omega_{X}h^2=0.03~(=25\%)$ & $\Omega_{\tilde{H}^0}h^2=0.06~(=50\%)$ &~\\
~ & $\tau_X=3\times 10^{27}$\,s & ~ & ~\\
\end{tabular}
\end{ruledtabular}
\caption{Our choice of couplings in the superpotential (defined in Eq \ref{eq:superpot}) and the resulting neutrino masses and dark matter properties. $m_a$ and $m_s$ denote the three active and sterile neutrino masses respectively. Along with these choices, we have set $\pv = 110$ PeV, $M_*=M_{\rm GUT} ~(= 10^{16}$ GeV), $\tan\beta=2$, and $T_{RH}=10^{10}$ GeV.}
\label{table:parameters}
\end{table}

The choice tan\,$\beta\!=\!2$ is compatible with the measured Higgs mass $m_h=125$ GeV with PeV scale superpartners. The cutoff scale $M_*$ is chosen to be the scale of grand unification $M_{GUT}~(= 10^{16}$ GeV), so the framework is expected to be embedded in a grand unified theory. With $\pv=110$ PeV, the specified values of $\zeta_{ij}$ and $\eta_i$ set the masses of sterile and active neutrinos ($m_s$ and $m_a$ respectively) via the see-saw mechanism. It can be seen that the entries are mostly $\mathcal{O}$(1), except for the third column of $\zeta_{ij}$ and the first entry in $\eta_i$, which are $\mathcal{O}(10^{-5})$; as mentioned in Section \ref{sterileneutrino}, this hierarchy is made inevitable by the need for $N_1$ to be at the keV scale and $N_{2,3}$ to be at the GeV scale for consistency with cosmology. It is worth noting that although $\mathcal{O}(10^{-5})$ seems unnaturally small, such a small number is already realized in nature in the form of the electron Yukawa coupling. With these choices, the lightest sterile neutrino $N_1$ has a mixing angle of sin$^2 (2\theta)\sim 4\times 10^{-10}$ and accounts for $\sim\!25\%$ of the dark matter abundance.

As mentioned earlier, the Higgsino is chosen to be the LSP. We take its mass to be $800$ GeV, so that it makes up about half of the dark matter. The scalar $X$ with mass $7$ PeV makes up the remaining fraction. Its abundance is controlled by the parameter $\alpha$ (for simplicity, we have taken a universal value $\alpha_i=\alpha$) and the reheat temperature $T_{RH}$ (which we have set to $10^{10}$ GeV). Likewise, its decay rate is controlled by the parameter $\lambda_2$. We see that the correct relic density and decay rate can be obtained for fairly reasonable values of these couplings.

\begin{figure}
\includegraphics[width=14cm]{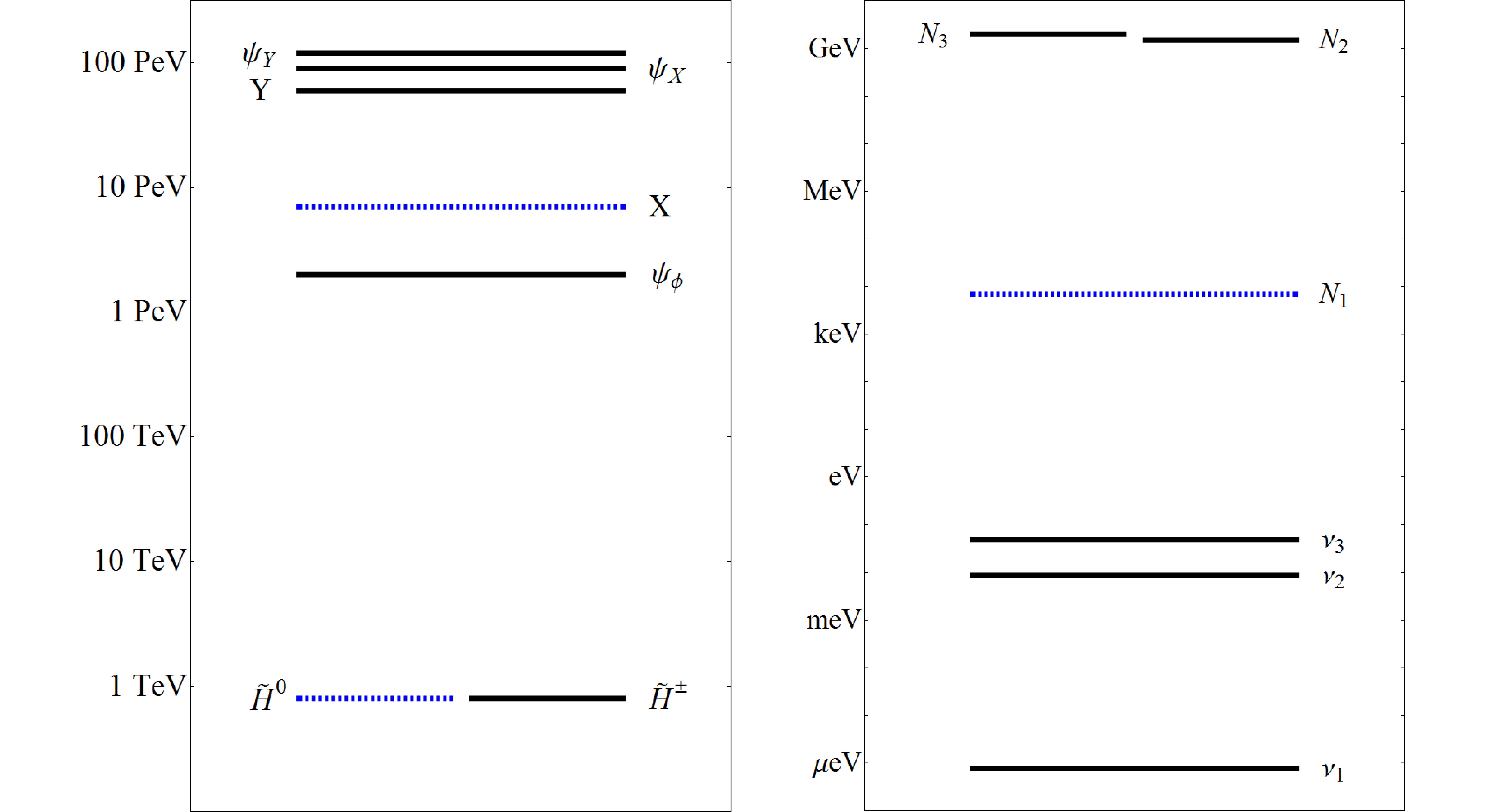}
\caption{The mass spectrum of our model. The left box displays the heavy PeV scale states as well as the TeV scale Higgsinos. The right box shows the active and sterile neutrino mass eigenstates. Particles that form some fraction of dark matter are denoted by dashed blue lines.}
\label{fig:spectrum}
\end{figure}

The spectrum of masses of various particles in the model relevant for this paper are plotted in Figure\,\ref{fig:spectrum}, with particles that contribute to dark matter highlighted in blue. We have an interesting scenario where dark matter is a mix of three components at very different scales: a 7 keV sterile neutrino, a stable 800 GeV Higgsino, and a $7$ PeV long-lived scalar.  For the dark matter distribution we adopt the Navarro-Frenk-White (NFW) profile \cite{Navarro:1995iw}, which was also found to be compatible with models of warm \cite{Lovell:2013ola} and mixed \cite{Harada:2014lma} dark matter, as well as the 3.5 keV line \cite{Boyarsky:2014jta}.

With these parameter choices, we now examine, in turn, the two signals of interest.

\subsection{PeV Neutrinos at IceCube}
\label{sec:icecube}

In this section we study the spectrum of neutrinos at IceCube from the decay of $X$ with the parameter choices above. For decaying dark matter, both galactic and extragalactic contributions are important. The galactic contribution to the neutrino flux $d\Phi_\alpha/dE_{\nu_\alpha}$, where $\alpha$ denotes the neutrino flavor, is given by \cite{Fong:2014bsa} \footnote{The coefficient in the first parenthesis differs from that in \cite{Fong:2014bsa} since we use an NFW profile instead of an Einasto profile.}
\begin{equation}
\left(\frac{d\Phi_{\alpha}}{dE_{\nu_\alpha}}\right)_\text{gal} =
\kappa \left( \frac{1.5\times 10^{-13}}{\rm cm^{2}\;sr\: s} \right)\left(\frac{10^{28} \rm  \, s}{\tau_{X}}\right) \left(\frac{\rm PeV}{m_X}\right) \frac{1}{N}\frac{dN_\alpha}{dE_{\nu_\alpha}}\,,
\end{equation}
where $\kappa=0.25$ is the fraction of dark matter made up by $X$, $1/N (dN_\alpha/dE_{\nu_\alpha})$ is the normalized neutrino energy spectrum from $X$ decay, and $\tau_X$ and $m_X$ are, respectively, the lifetime and mass of $X$. Likewise, the extragalactic contribution is \cite{Fong:2014bsa, Ibarra:2007wg}
\begin{equation}
\left(\frac{d\Phi_{\alpha}}{dE_{\nu_\alpha}}\right)_\text{ex-gal} =
\kappa \left(\frac{2.5\times10^{-13}}{\rm cm^{2} \, sr\, s}\right) \left(\frac{10^{28} \rm  \, s}{\tau_{X}}\right) \left(\frac{\rm PeV}{m_X}\right) \int_{1}^{\infty}dy\frac{dN_\alpha}{N \,
  d(E_{\nu_\alpha}y)}\frac{y^{-3/2}}{\sqrt{1+(\Omega_{\Lambda}/\Omega_{M}) \,
    y^{-3}}}\, ,
\end{equation}
where $\Omega_M = 0.315$, $\Omega_\Lambda = 0.685$, and $y=1+z$, where $z$ is the redshift. The expected number of neutrino events at IceCube in the energy bin between $E_1$ and $E_2$ is given by
\begin{equation}
N(E_1,E_2)=4\pi\, T \, \sum_{\alpha} \, \int_{E_1}^{E_2} dE_{\nu_\alpha} \,  A^{\alpha}_{\rm
  eff}(E_\nu) \left( \frac{d\Phi_{\alpha}}{dE_{\nu_\alpha}} \right)_\text{gal\,+\,ex-gal}\, ,
\label{eq:neutrinorate}
\end{equation}
where $T=988$ days is the total exposure time \cite{Aartsen:2014gkd} and $A^{\alpha}_{\rm eff}(E_\nu)$ is the IceCube effective area for neutrino flavor $\alpha$, taken from \cite{Aartsen:2013bka}.

\begin{figure}
\includegraphics[width=13cm]{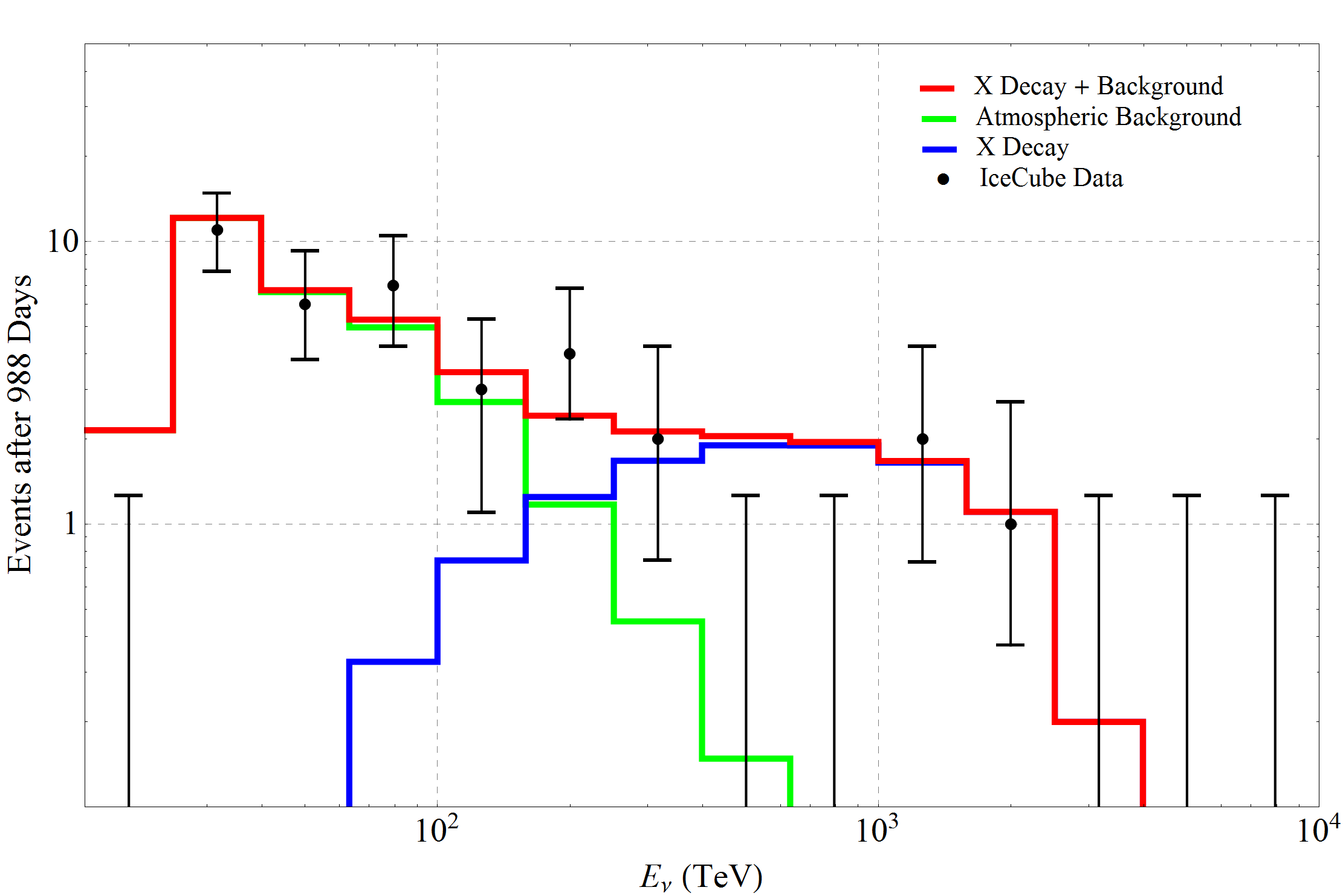}
\caption{The expected number of neutrino events at IceCube from the decay of $X$ (blue), together with IceCube data points and error bars (black) and atmospheric background (green) from \cite{Aartsen:2014gkd}, and the total number of signal+background events (red).}
\label{fig:icecube}
\end{figure}

We entered the model described in Section \ref{sec:model} into PYTHIA 8.2 \cite{Sjostrand:2014zea} and generated the neutrino spectrum, $dN_\alpha/dE_{\nu_\alpha}$ from $X$ decay following the decay chain specified in Section \ref{sec:scalarx}. The various decay channels and rates for the $\mathcal{O}$(keV-GeV) mass sterile neutrinos are listed in the appendix of \cite{Gorbunov:2007ak}. The number of neutrino events in each IceCube bin thus predicted following the above procedure is shown in Figure\,\ref{fig:icecube} in blue. The IceCube data points (black dots), error bars, and background (green) are taken from \cite{Aartsen:2014gkd}. The background events come from cosmic ray air showers, which produce $\pi/K$, and expected charmed mesons, which decay to muons and neutrinos; the background reported here is the sum of the average background from atmospheric muons and neutrinos, plus the 90\% confidence level upper limit for neutrinos from charmed decays \cite{Aartsen:2014gkd}. The total number of events expected (signal+background) for the reported exposure is shown in red.
%The mass of the long lived scalar $m_X$ sets the energy at which the decay spectrum peaks, while the combination $\frac{\kappa}{m_X \tau_X}$ sets the normalization.

From the plot, it is clear that the measurements in the lower energy bins are completely consistent with atmospheric background, whereas the higher energy bins (200 TeV and above) show a clear deviation from what is expected from background only, indicating an excess of neutrino events that requires some explanation. This excess has four salient features:

\indent(i) an excess in two bins at $200-400$ TeV,\\
\indent(ii) no events in the two bins covering $500$ TeV - $1$ PeV,\\
\indent(iii) three PeV scale events, two at 1 PeV and one at 2 PeV, and \\
\indent(iv) no events above 2 PeV. \\
An additional astrophysical power-law contribution could address (i), but clearly cannot explain the three neutrinos with PeV energies (iii) \cite{Fong:2014bsa}.

Models of PeV dark matter in the literature claim to be able to explain these features by employing a two-body decay channel that includes a neutrino to give (iii) as well as a secondary decay channel that gives softer neutrinos to explain (i) \cite{Fong:2014bsa, Ema:2013nda, Ema:2014ufa, Feldstein:2013kka, Esmaili:2013gha, Esmaili:2014rma, Bai:2013nga, Bhattacharya:2014vwa,Higaki:2014dwa,Bhattacharya:2014yha,Rott:2014kfa}. In contrast, our model does not have a direct two-body decay channel into neutrinos, hence there is no sharp feature that peaks at PeV energies. Instead, since neutrinos are produced via two- or three- step decay chains, each involving multiple decay products, the neutrino spectrum is essentially flat, with a dropoff at approximately half the dark matter mass (blue curve in Figure\,\ref{fig:icecube}). The flat spectrum allows us to generate a signal contribution that satisfactorily addresses both (i) and (iii), but at the cost of disagreeing with (ii). However, since less than 2 events are predicted in each of the two bins in (ii), the disagreement is not too fatal and can be attributed to a possible downward fluctuation of the signal. Hence the model predicts that events should appear in these bins when more data is collected. Finally, although the DM mass is chosen to have the spectrum drop off after the 2 PeV bin, hence explaining (iv), we note that somewhat heavier masses would still predict less than 1 event in each of the bins higher than 2 PeV and hence could remain compatible with the current data. Consequently, possible future measurements of events at energies higher than 2 PeV need not necessarily be incompatible with a dark matter explanation.

\vskip 0.5cm
\noindent\textit{Additional Constraints}
\vskip 0.3cm

In addition to neutrinos, the dark matter decay products can also give visible signatures in other channels. The most important of these is gamma rays. The $e^\pm$ from DM decays produce energetic photons due to inverse Compton scattering and synchrotron radiation. Likewise, since the Universe is opaque to gamma rays with energies above a TeV, high energy gamma rays produce $e^\pm$ pairs through interactions with the interstellar radiation field. Such cascades from high energy products from DM decay therefore produce a population of gamma rays between $\mathcal{O}$(1) GeV and $\mathcal{O}$(100) GeV \cite{Semikoz:2003wv}. Following \cite{Esmaili:2014rma}, we verify the compatibility of the DM decay process with gamma-ray bounds from the Fermi-LAT measurement of the isotropic diffuse gamma-ray background \cite{Ackermann:2014usa} by considering the integrated energy density $\omega_\gamma$ measured by the Fermi-LAT between $E_1 \sim \mathcal{O}(1)$ GeV and $E_2 \sim \mathcal{O}(100)$ GeV \cite{Esmaili:2014rma} 
\begin{equation}
\label{fermilimit}
\omega_\gamma = \frac{4 \pi}{c} \int_{E_1}^{E_2} E_\gamma \frac{d\Phi_\gamma}{dE_\gamma} dE_\gamma \approx 4.4 \times 10^{-7} \,{\rm eV/cm^3}\,.
\end{equation}
The total energy density in photons and $e^\pm$ from the decay of $X$, using the output from PYTHIA, is calculated to be $\sim\!4 \times 10^{-9}$ eV/cm$^3$. This can be interpreted as the maximum amount of energy that can be deposited in the diffuse gamma-ray background; since this is well below the energy density measured by Fermi quoted above, we conclude that the DM decay process is not in tension with gamma-ray measurements \footnote{See \cite{Murase:2015gea} for more stringent (but model-dependent) constraints from analyzing the spectrum rather than the integrated energy.}. Likewise, we note that since the DM decays exclusively to leptons, it is unlikely to be in tension with the recent \AMS\ measurements of the $\bar{p}/p$ ratio \cite{AMS} (from extrapolating Figure\,5 of \cite{Giesen:2015ufa}, which shows the bound on the lifetime of a DM particle decaying to $b\bar{b}$ from antiproton constraints, to $\mathcal{O}(10)$ PeV, it can be seen that even a hadronically decaying particle with a lifetime of $\sim\!10^{27}$s would be well within the antiproton limits).

%%%%%%%%%%%%%%%%%%%%%%%%%%%%%%%%
\subsection{3.5 keV X-ray Signal}
\label{sec:xray}
%%%%%%%%%%%%%%%%%%%%%%%%%%%%%%%%

Several papers have interpreted this signal as the decay of a $\sim\!7$ keV sterile neutrino dark matter component \cite{Bulbul:2014sua, Boyarsky:2014jta, Abazajian:2014gza, Merle:2014xpa, Haba:2014taa, Rodejohann:2014eka, Abada:2014zra, Ishida:2014fra, Robinson:2014bma, Chakraborty:2014tma, Adulpravitchai:2014xna, Patra:2014pga, Frigerio:2014ifa, Tsuyuki:2014aia, Kang:2014mea, Harada:2014lma, Ishida:2014dlp, Bezrukov:2014nza}.  A keV scale sterile neutrino can be produced through the Dodelson-Widrow (DW) mechanism \cite{Dodelson:1993je} with an approximate relic abundance
\beq
\Omega h^2\sim 0.07 \left(\frac{\sin^2 2\theta}{10^{-9}}\right)\left(\frac{m_s}{7 ~\text{keV}}\right)^2.
\eeq
This is a consequence of non-resonant oscillation due to the mixing between the active and sterile sectors. However,  sterile neutrinos produced through the DW mechanism accounting for all of dark matter has now been robustly ruled out for all masses based on constraints from X-ray bounds  \cite{Boyarsky:2006fg,Boyarsky:2006ag, Watson:2006qb,Boyarsky:2005us,Boyarsky:2007ay,Boyarsky:2007ge,Watson:2011dw,Horiuchi:2013noa} and Lyman-alpha data \cite{Seljak:2006qw, Asaka:2006nq, Boyarsky:2008xj}; see \cite{Horiuchi:2013noa} for a summary.

This problem can be evaded in several ways. A lepton asymmetry in the early Universe can lead to resonant production of sterile neutrinos, resulting in a colder distribution that can evade Lyman-alpha bounds; this is known as the Shi-Fuller mechanism \cite{Shi:1998km}. Another approach is to consider sterile neutrino production from the decays of a singlet Higgs boson from an extended Higgs sector \cite{Kusenko:2006rh,Petraki:2007gq}, which can produce all of dark matter from the freeze-in mechanism \cite{Hall:2009bx} (for a discussion in the context of the 3.5 keV line, see \cite{Merle:2014xpa}). 

Likewise, the constraints are avoided if sterile neutrinos make up only a fraction of dark matter. While it is trivial to rescale the X-ray constraints to account for a smaller fraction of dark matter, a reinterpretation of the Lyman-alpha constraint is not straightforward and requires numerical simulations. An analysis in \cite{Boyarsky:2008xj}, for instance, showed that $m_s \geq 5$ keV warm component constituting $\leq 60 \%$ of the total dark matter abundance, where the rest of the dark matter is made up of a cold component, is consistent with all constraints \cite{Boyarsky:2009ix}; a follow-up study \cite{Boyarsky:2008mt} found similar results. In a warm plus cold dark matter setup, \cite{Harada:2014lma} found that the 3.5 keV X-ray signal could be explained with a 7 keV sterile neutrino produced from DW that made up $10-60 \%$ of dark matter. We construct our theory to map on to one of the benchmark points in \cite{Harada:2014lma}, which were shown to satisfy the relevant constraints. In \cite{Harada:2014lma}, the signal was found to be compatible with a non-resonantly produced $7$ keV sterile neutrino with mixing angle sin$^2 (2\theta)\sim 4\times 10^{-10}$, making up $\sim 25 \%$ of dark matter; we reproduce these values with our choice of parameters in Table \ref{table:parameters}, thereby incorporating the $3.5$ keV X-ray line in our framework.

A mixture of warm and cold components for dark matter offers several advantages. Ref. \cite{Harada:2014lma} found that this scenario was still compatible with an NFW profile and could resolve the missing satellite problem \cite{Moore:1999nt, Kravtsov:2009gi}. In addition, we note that the larger free-streaming length of the non-resonantly produced sterile neutrino, which is a warm dark matter component, could result in it being underabundant in dwarf galaxies (which would be made up mostly of the cold component), possibly providing an explanation of the non-observation of the signal in the stacked analysis of dwarf spheroidals presented in \cite{Malyshev:2014xqa}; establishing this would, however, require detailed numerical simulations that are beyond the scope of this paper.

\section{Summary}

In this paper we have attempted to incorporate two recent potential hints of new physics, the PeV neutrinos at IceCube and the $3.5$ keV X-ray line, into a broader, independently motivated framework of a PeV scale supersymmetric neutrino sector.
\begin{itemize}
\item The right handed sterile neutrinos are expected to be charged under some new symmetry (e.g., $U(1)'$) beyond the SM gauge group, which enables light masses at phenomenologically interesting scales (keV-GeV). In order to successfully realize the desired active and sterile neutrino masses without unnaturally small parameters, this symmetry must be broken by a PeV scale vev, which corresponds to a desired scale in some approaches to supersymmetry breaking.
\item The lightest sterile neutrino can have a mass of $7$ keV and be produced non-resonantly through the DW mechanism to form a fraction of dark matter; its decay can explain the $3.5$ keV X-ray line.
\item The neutrino sector can be extended to include additional fields with similar operators to the ones that give rise to neutrino masses. This can give a PeV scale dark matter component whose relic abundance is set by freeze-in processes, and an appropriate charge under the $U(1)'$ symmetry makes it long-lived and gives a decay spectrum into neutrinos consistent with the high energy events observed at IceCube.
\end{itemize}

Although we have worked in a setup where dark matter consists of $25\%$ sterile neutrinos, $25 \%$ PeV scalar $X$, and $50\%$ Higgsino LSP, it should be clear that the dark matter composition can be completely different from this admixture, given that the PeV supersymmetric sector is likely far more complicated, and neither of the two signals are therefore necessary ingredients of the extended neutrino sector we have discussed. Since what is presented here only serves as an illustrative proof of principle and can incorporate several variations, we did not attempt to perform detailed fits or delve into model-building details beyond what was necessary, although such exercises might be warranted once more data is collected in the future, and can shed light on the underlying model. These should not distract from the most central ideas of the paper that these two signals, at such different energy scales, fit rather naturally into a broader particle physics framework and could be the first indications of a rich, hitherto unexplored neutrino sector.

\medskip
\textit{Acknowledgements: }The authors are supported in part by the DoE under grants DE-SC0007859 and DE-SC0011719. BS also thanks the Center for Theoretical Underground Physics and Related Areas (CETUP*), where part of this work was completed, for its hospitality.
%%%%%%%%%%%%%%%%%%%%%%%%%%%%%%%%%%%
\bibliography{pevkevbib}

\end{document}